\documentclass[prd,twocolumn,nofootinbib]{revtex4}
\usepackage{graphicx} 
\usepackage{lipsum} 
\usepackage{amsmath}
\usepackage{subcaption}
\usepackage{siunitx}
\usepackage{svg}
\usepackage{url}

\begin{document}

\title{Demonstrating the velocity response of a table-top EPR Speedmeter}
\author{S.L. Kranzhoff $^{1,2}$, S.L. Danilishin$^{1,2}$, S. Steinlechner$^{1,2}$, M. Vardaro$^{1,2}$, T. Zhang$^{3}$ and S. Hild$^{1,2}$}
\affiliation{$^{1}$Maastricht University, Department of Gravitational Waves and Fundamental Physics, 6200MD Maastricht, the Netherlands}
\affiliation{$^{2}$Nikhef, Science Park 105, 1098 XG Amsterdam, the Netherlands}
\affiliation{$^{3}$Institute for Gravitational Wave Astronomy, School of Physics and Astronomy, University of Birmingham, Birmingham B15 2TT, United Kingdom}

\begin{abstract}
The sensitivity of gravitational-wave interferometers is fundamentally limited by quantum noise, as dictated by the Heisenberg uncertainty principle, due to their continuous position measurement of the end mirrors. Speedmeter configurations, which measure mirror velocity rather than position, have been proposed as a means to suppress quantum back-action noise, but practical implementations remain at an early stage. In this work, we present a table-top realisation of the Einstein-Podolsky-Rosen (EPR) Speedmeter concept, employing an optical readout scheme based on two orthogonal polarisation modes that probe the interferometer with different effective bandwidths. Using a triangular cavity, we demonstrate that the differential optical response between the linear p- and s-polarised modes exhibits a speed-like frequency dependence: vanishing at DC and increasing linearly with signal frequency, up to the bandwidth of the slower mode. With this we show that an optical system equivalent to the EPR Speedmeter indeed performs a velocity readout of the end mirror.
\end{abstract}
\maketitle

\section{Introduction}
Second-generation kilometer-scale gravitational-wave interferometers such as Advanced LIGO~\cite{aLIGO2015} and Advanced Virgo~\cite{aVirgo2015} have successfully detected signals from compact binary coalescences, including merging black holes and neutron stars, during their first three joint observing runs~\cite{GWTC2019,GWTC2024,GWTC2023}. In the ongoing fourth observing run (O4), the cryogenic underground detector KAGRA in Japan~\cite{Akutsu2021} has joined the global detector network and more than 200 candidate events have been reported, offering among other things valuable insights into stellar evolution and the population of massive compact objects in the Universe. Although current detectors have not yet reached design sensitivity across the full detection band~\cite{aLIGO-O4performance2024}, ongoing upgrades and the development of next-generation observatories such as the Einstein Telescope (ET)~\cite{ET2010,ETscMaggiore2020,ETblue2025} are expected to push quantum noise to the forefront as the dominant sensitivity limitation across much of the detection bandwidth~\cite{Martynov2016}.

Quantum noise arises from fundamental vacuum fluctuations entering the interferometer via its dark port, contributing in two uncorrelated ways~\cite{Caves1981, Braginsky1967, DanilishinLivRev2012}. First, quantum phase fluctuations manifest as shot noise in the photodetector readout, dominant at high frequencies. Second, quantum amplitude fluctuations couple to the carrier light in the interferometer arms, generating a fluctuating radiation pressure force on the mirrors—so-called back-action noise—which scales inversely with the square of signal frequency, $\sim\Omega^{-2}$, due to the mirrors acting as free test masses. These two noise terms trade off against each other: increasing circulating optical power reduces shot noise but enhances back-action. The frequency at which both contributions become equal defines the Standard Quantum Limit (SQL), beyond which position-measurement interferometers with uncorrelated shot and back-action noise cannot surpass without altering their quantum measurement scheme.

Modern gravitational-wave detectors already employ squeezed vacuum states to reduce quantum noise~\cite{Vahlbruch2010, GEOsqz2011, LIGOsqz2019, Virgosqz2019}. This technique exploits quantum correlations to suppress fluctuations in one quadrature (e.g., phase) at the cost of enhancing those in the conjugate quadrature (e.g., amplitude), in accordance with the Heisenberg uncertainty principle~\cite{Walls1983, Schnabel2017, Danilishin2019}. Simultaneous broadband suppression of both noise types requires frequency-dependent squeezing, typically achieved using large-scale filter cavities, which has recently been demonstrated in full-scale detectors~\cite{Ganapathy2023, AcerneseFDS2023}.

An alternative quantum noise mitigation strategy is offered by speedmeter interferometer configurations, first proposed in~\cite{Braginsky1990}. In contrast to position meters, speedmeters aim to measure the momentum of the test masses, which is a quantum non-demolition (QND) observable. Such observables satisfy the commutation relation $\left[\hat{o}(t), \hat{o}(t')\right]=0$ for all times $t, t'$, and can therefore, in principle, be measured continuously without introducing quantum back-action~\cite{Braginsky1980}. A speedmeter achieves this by performing two coherent position measurements with opposite sign in the interaction Hamiltonian and a short delay $\tau$, so that the readout becomes proportional to the average velocity, $\phi_{\text{out}}\propto x(t+\tau)-x(t)\simeq\overline{v}\tau$, and the radiation pressure forces imparted by each measurement partly cancel. Note that velocity is not strictly proportional to the momentum of the free test mass when it is coupled to a meter, so that a speedmeter achieves a significant reduction but not a complete cancellation of back-action noise~\cite{Danilishin2019}. 

Multiple realisations of the speedmeter concept have been proposed over the years~\cite{Braginsky2000, Purdue2002, Chen2003, Wade2012, Wang2013, Huttner2017, Knyazev2018, Danilishin2019}, but none have yet reached the technological maturity required for large-scale implementation. In this paper, we present an experimental demonstration of a table-top Einstein-Podolsky-Rosen (EPR) Speedmeter, as proposed in~\cite{Knyazev2018}. This approach belongs to the class of polarisation-based speedmeters, which employ two orthogonal polarisation modes to perform two independent measurements of the test mass position. A key advantage of this scheme is that it does not require major modifications to the core interferometer topology; only additional optical elements at the interferometer output are needed to convert position readout into speed readout.

We implement the EPR Speedmeter using a triangular optical cavity with polarisation-dependent bandwidths resulting in different frequency responses acquired by p-polarised and s-polarised modes. We show that the differential optical response between these modes exhibits a speed-like frequency-dependence: vanishing at DC and increasing linearly with signal frequency, up to the bandwidth of the slower mode. \par 
The structure of the paper is as follows. Section~\ref{sec:idea} outlines the theoretical concept of the EPR speedmeter and its equivalence to a triangular cavity. Section~\ref{sec:meas} describes the experimental setup and measurement procedure. Section~\ref{sec:result} presents the results, and Section~\ref{sec:discussion} provides a discussion and outlook.

\section{The triangular cavity as an EPR Speedmeter}
\label{sec:idea}
\begin{figure}[h]
    \centering
    \includegraphics[width=\linewidth]{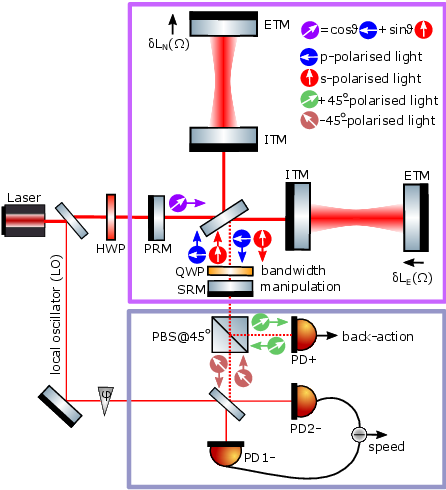}
    \caption{Optical layout of the EPR Speedmeter as gravitational-wave interferometer adapted from \cite{Knyazev2018}.}
    \label{fig:eprsm-scheme}
\end{figure}

The EPR Speedmeter shown Fig.~\ref{fig:eprsm-scheme} features a dual-recycled Michelson interferometer with Fabry-Perot arm cavities (purple box) pumped with linearly polarised laser light at polarisation angle $\vartheta$ set by the orientation of the half-wave plate (HWP) at the interferometer input. The quarter-wave plate (QWP) at the interferometer output is oriented such that it introduces a phase retardation of $\pi/2$ between p-polarised (blue) and s-polarised (red) components of the light used to probe the differential arm length change $\delta L_{\text{darm}}(\Omega)=\delta L_{\text{N}}(\Omega)-\delta L_{\text{E}}(\Omega)$ caused by a passing gravitational wave at frequency $\Omega$. The position of the signal-recycling mirror (SRM) is then tuned such that one polarisation mode (e.g. s-pol) is resonant in the signal-recycling cavity formed by the SRM and the input test masses (ITM) of the two Fabry-Perot arm cavities, while the other one (e.g. p-pol) experiences the resonant sideband extraction configuration of the signal-recycling cavity resulting in a much larger interferometer bandwidth $\gamma_{\text{p}}\gg\gamma_{\text{s}}$ in accordance with the so-called scaling law~\cite{Buonnano2003}. The interferometer bandwidth is inversely proportional to the average storage time of light in the interferometer and can therefore be understood as the rate at which position information leaks out of the interferometer. Due to the different bandwidths for the two polarisation modes, a recombination at a beamsplitter effectively combines position information from different times and therefore results in a speed measurement.

In case of the EPR Speedmeter, a polarising beamsplitter (PBS) rotated by $45^{\circ}$ acts as 50:50 beamsplitter for p-polarised and s-polarised light splitting the light into $\pm 45^{\circ}$ linearly polarised beams where the difference channel (transmission) carries speed information that can be extracted by choosing a suitable local oscillator phase $\varphi$. The vacuum entering through the sum port of the PBS creates additional back-action noise but can be removed in post-processing by measuring the amplitude quadrature of the sum channel (PD+). This is possible since the position meter outputs represented by the two polarisation modes get entangled when combined on the PBS, which motivated the name of this speedmeter configuration after the famous Gedankenexperiment by Einstein, Podolsky and Rosen.

\begin{figure}[h]
    \centering
    \includegraphics[width=\linewidth]{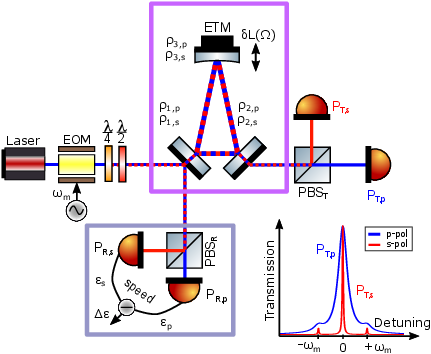}
    \caption{Conceptual design of the EPR Speedmeter proof-of-principle experiment with a triangular cavity representing a system with two different bandwidths for two linear polarisation modes (purple box) and a speed readout by means of the error signal difference $\Delta\epsilon=\epsilon_{\text{p}}-\epsilon_{\text{s}}$ (grey box).}
    \label{fig:triCav-concept}
\end{figure}

In a nutshell, the EPR Speedmeter achieves a speed measurement by using two polarisation modes with different bandwidths $\gamma_{\text{p}}\gg\gamma_{\text{s}}$ which results in different AC responses of the interferometer. The differential response vanishes at DC and is to leading order proportional to the signal frequency $\Omega$, therefore, it represents speed. A triangular ring cavity, see Fig.~\ref{fig:triCav-concept}, similarly realises a system with two different bandwidths for p-polarised and s-polarised light, since the transmissivity of the input mirror is in general different for the two orthogonal polarisation modes under non-normal incidence as described by Fresnel relations.

The frequency-dependent reflectivity of a triangular ring cavity for each polarisation mode $k=$p, s can be derived as
\begin{align}
\label{eq:triCav-ref}
    \mathcal{R}_{k}\left(\Omega\right)=\frac{\rho_{1,k}-\rho_{2,k}\rho_{3,k}e^{\mathrm{i}\phi_{\text{rt}}(\Omega)}}{1-\rho_{1,k}\rho_{2,k}\rho_{3,k}e^{\mathrm{i}\phi_{\text{rt}}(\Omega)}}\,,
\end{align}
where $\rho_{j,k}=\sqrt{R_{j,k}}$ is the amplitude reflectivity of each mirror $j$ derived from the power reflectivity $R_{j,k}$ specified by the optics supplier. For lossless mirrors, reflectivity and transmissivity are connected via $\rho_{j,k}=\sqrt{1-\tau_{j,k}^2}$ due to energy conservation. Assuming a high reflectivity for the piezo-driven end mirror $\rho_{3,k}\approx 1$ for both polarisations, we define the coupling rates of the two input mirrors $\gamma_{j,k}$ and the complex cavity pole function $\ell_{k}(\Omega)$ as
\begin{align}
    \gamma_{j,k}=\frac{\tau^{2}_{j,k}}{2\mathcal{T}_{\text{rt}}}\,,\hspace{0.5cm} \ell_{k}(\Omega)=\gamma_{1,k}+\gamma_{2,k}-\mathrm{i}\Omega\,,
\end{align}
with $\mathcal{T}_{\text{rt}}=L_{\text{rt}}/c$ the travel time of light for a single pass through the cavity with round-trip length $L_{\text{rt}}$. Using the single-mode approximation with assumptions $\Omega\mathcal{T}_{\text{rt}}\ll1$ and $\tau_{j,k}^{2}\ll1$ combined with operation close to resonance, $\phi_{\text{rt}}=2\pi n+\Omega\mathcal{T}_{\text{rt}}$ ($n$ integer), one can show that 
\begin{align}
    \mathcal{R}_{k}(\Omega)\approx\frac{\gamma_{2,k}-\gamma_{1,k}+\mathrm{i}\Omega}{\gamma_{1,k}+\gamma_{2,k}-\mathrm{i}\Omega}\,.
\end{align}
A mirror motion $\delta L(\Omega)$ at frequency $\Omega$ (signal) creates symmetric sidebands at $\omega_{0}\pm\Omega$ around pump frequency $\omega_{0}=k_{0}c$ of the laser, which we treat as DC at the cavity operating point. These sidebands leak through the input mirror and show up in the reflected field,
\begin{align}
    E_{\text{ref},k}(\Omega)&\approx\frac{2\mathrm{i}k_{0}\sqrt{\gamma_{1,k}}}{\ell_{k}(\Omega)\sqrt{\mathcal{T}_{\text{rt}}}}E_{\text{in},k}\cdot\delta L(\Omega)\\
    &=H_{\text{opt},k}(\Omega)\cdot\delta L(\Omega)\,,
\end{align}
with $E_{\text{in},k}$ input fields to the cavity calculated from the optical power via $P_{k}=\hbar\omega_{0}E_{\text{in},k}^{2}$.

The optomechanical transfer function $H_{\text{opt},k}$ describes how motion of the mirror couples to the electric field of each polarisation $k$ and can be extracted using the error signal $\epsilon_{k}\propto H_{\text{opt},k}(\Omega)\cdot\delta L(\Omega)$ of the well-established Pound-Drever-Hall (PDH) technique \cite{Drever1983}. The input field $E_{\text{in},k}$ contains frequency components of the carrier $\omega_{0}$ and symmetric sidebands at $\omega_{0}\pm\omega_{\text{m}}$ from phase modulation with the electro-optic modulator (EOM) at frequency $\omega_{\text{m}}$. The EOM modulation frequency is chosen large enough such that $\mathcal{R}_{k}(\omega_0\pm\omega_{\text{m}})\approx 1$, which means that the sidebands are reflected from the cavity and therefore not affected by mirror motion, which creates new sidebands at $\omega_{0}\pm\Omega$. The optical power reflected from the cavity is separated into p-polarised and s-polarised components by means of a PBS (see Fig.~\ref{fig:triCav-concept}) to monitor the reflected power $P_{\text{R},k}$ for both polarisations independently with photodetectors. The PDH error signals $\epsilon_{k}$ for both polarisations are generated by mixing the measured reflection signal with a reference at EOM modulation frequency $\omega_{\text{m}}$ and low-pass filtering. This generates signals at $\omega_{\text{m}}\pm\Omega$, one of which beats down to a demodulated signal at $\Omega$. 

For a suitable mirror motion amplitude, the difference between the error signals of the two polarisations is
\begin{align}
\label{eq:diff-resp}
    \frac{\Delta\epsilon(\Omega)}{\delta L(\Omega)}&\propto\frac{2\mathrm{i}k_{0}}{\sqrt{\mathcal{T}_{\text{rt}}}}\left(\frac{E_{\text{in,p}}\sqrt{\gamma_{1\text{,p}}}}{\ell_{\text{p}}(\Omega)}-\frac{E_{\text{in,s}}\sqrt{\gamma_{1\text{,s}}}}{\ell_{\text{s}}(\Omega)}\right)\\
    &\overset{(*)}{=}\frac{2\mathrm{i}k_{0}}{\sqrt{\mathcal{T}_{\text{rt}}}}\frac{E_{\text{in,p}}}{\sqrt{\gamma_{\text{1,p}}}}\frac{\mathrm{i}\Omega(\gamma_{1\text{,s}}-\gamma_{1\text{,p}})}{\ell_{\text{p}}(\Omega)\ell_{\text{s}}(\Omega)}\propto\mathrm{i}\Omega
    \,,
\end{align}
where the last equality sign is valid for a critically coupled cavity $\gamma_{1,j}=\gamma_{2,j}$ if the tuning condition $(\ast)\, E_{\text{in,p}}/\sqrt{\gamma_{1\text{,p}}}=E_{\text{in,s}}/\sqrt{\gamma_{1\text{,s}}}$ is fulfilled, equivalent to the tuning condition formulated for the EPR Speedmeter~\cite{Knyazev2018,Danilishin2019}. The above expression is linear in frequency and the system behaves like a speedmeter for $\Omega<\min(\gamma_{1\text{,p}}, \gamma_{1\text{,s}})$ provided that $\gamma_{1\text{,p}}\neq\gamma_{1\text{,s}}$. Considering only optical parameters, the tuning condition is fulfilled by choosing the input polarisation angle $\vartheta=\arctan(\gamma_{1\text{,s}}/\gamma_{1\text{,p}})$. Gain differences from other sources such as photodetector response and readout electronics have been omitted in Eq.~\ref{eq:diff-resp}, but need to be taken into account in the real experiment by proper calibration as detailed in Sec.~\ref{sec:meas}. 

In the following, the experimental realisation of such a system as a table-top experiment is discussed. We demonstrate a classical speed-like response of the triangular cavity representation of the EPR Speedmeter. Note that we do not show back-action suppression in this paper, as this would require a quantum-noise limited system including, among other things, a vacuum system, monolithic suspensions and a high-power laser with external frequency stabilisation.

\section{Experimental realisation}
\subsection{Setup and control}
\label{sec:meas}
\begin{figure*}
    \centering
    \includegraphics[width=\textwidth]{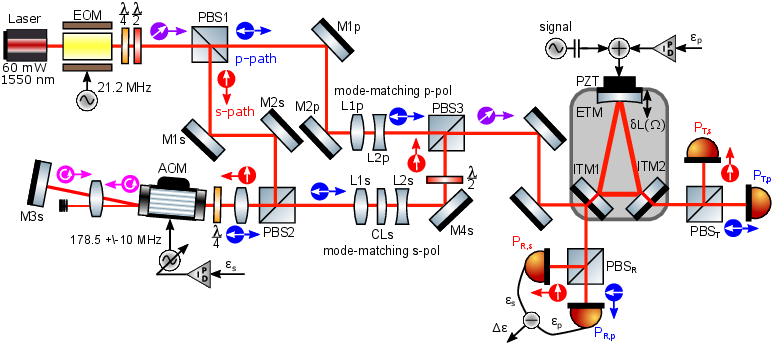}
    \caption{Practical realisation of the triangular cavity version of the EPR Speedmeter on an optical bench. Laser light enters from the left and both polarisation modes are prepared separately for injection into the optical cavity. The cavity spacer has three mirrors clamped to it, with the input/output couplers (ITM1, ITM2) featuring different transmissivities for the two polarisation modes and the piezo-driven, highly reflective end mirror (ETM) used for locking the cavity length to the p-polarisation mode and for signal injection. The speed-like response is measured by recording the differential demodulated error signal $\Delta\epsilon$ in reflection. The transmission signals are used for lock acquisition and optical characterisation.}
    \label{fig:triCav_full-scheme}
\end{figure*}

The optical layout in Fig.~\ref{fig:triCav_full-scheme} shows the realisation of the idea described in the previous section and highlighted in Fig.~\ref{fig:triCav-concept}. As input to the experiment, 60\,mW of laser power at wavelength 1550\,nm was used. The input polarisation was chosen to be linear by adjusting the QWP and along a polarisation angle by adjusting the HWP, such that signals of similar amplitude within range of the two reflection photodetectors were measured. Additionally, phase-modulated sidebands were imprinted by a resonant EOM at 21.2\,MHz for both polarisation modes to enable PDH lock and signal extraction from the PDH error signal.

For a triangular cavity and any other cavity with an odd number of mirrors, the resonance condition for horizontal modes is shifted by half a free spectral range (FSR) compared to vertical modes. In Jones formalism, this is represented by a sign flip acquired by the s-polarised electric field component at each reflection, which for simplicity is not taken into account in Eq.~\ref{eq:triCav-ref}. In order to have the fundamental mode of both polarisations resonant in the cavity at the same time, one of the polarisation modes needs to be frequency-shifted by FSR/2 with respect to the other. For this purpose, the two polarisation components are split by PBS1 and prepared separately before being recombined by PBS3 and injected into the cavity, see Fig.~\ref{fig:triCav_full-scheme}. 

By design, the cavity spacer used for the experiment had a round-trip length of 0.42\,m, which corresponds to $\mathrm{FSR}=c/L_{\text{rt}}\approx 714\,$MHz. An accousto-optical modulator (AOM) with nominal driving frequency 178.5\,MHz was used in double-pass configuration to achieve a total frequency shift of $\mathrm{FSR}/2\approx 357\,$MHz. The AOM allows for frequency modulation of $\pm 10\,$MHz by means of an applied DC voltage in the range $\pm 2\,\text{V}_{\text{dc}}$, which is used for PDH lock of the s-polarisation mode accounting for FSR differences from manufacturing tolerances and rubber rings between spacer, piezo and end mirror. 

The direction of the AOM output beam depends on the applied frequency shift. Operation in double-pass configuration ensures that the beam alignment to the cavity is independent of the frequency shift applied by the AOM, since the mirror M3s retro-reflects the beam independently of its direction. However, the double-pass configuration has the disadvantage of introducing large losses, since only up to $55\%$ of the input power are coupled into the first order diffracted beam. The orientation of the HWP at the input was therefore chosen such that 55\,mW of the original 60\,mW input power were sent to the s-path. A double-pass through the QWP infront of the AOM turns s-polarised light into p-polarised light which ensures separation of input and output beams from the AOM by means of PBS2. Additionally, the AOM introduces an astigmatism quantified by the ratio of beam waist sizes $w_{0i}$ in two orthogonal directions ($i$=x,y) $w_{0\text{x}}/w_{0\text{y}}\approx 0.54$, which needs to be compensated by a cylindrical lens (CLs) in order to avoid further losses from mode-mismatch to the cavity. 

The triangular cavity itself had flat mirrors with identical specifications for ITM1 and ITM2 with power transmissivities $T_{\text{i,p}}\approx 4.3\%$ for p-polarised light and $T_{\text{i,s}}\approx 0.35\%$ for s-polarised light, respectively, at an angle of incidence $44^{\circ}$. The piezo-driven cavity end mirror (ETM) was chosen to be highly reflective for both polarisations, $T_{\text{p,s}}=2\,$ppm, and with radius of curvature 1\,m. In this configuration, the fundamental TEM$_{00}$ mode resonant in the triangular cavity had a beam waist of $448\,\mu$m located in the centre between the two ITMs. The positions and focal lengths of the mode-matching telescopes for p-polarised light (lenses L1p, L2p) and s-polarised light (lenses L1s, L2s) were chosen and adjusted to match this condition independently for both polarisations. 

The power transmissivities of the two ITMs lead to effective measured bandwidths of $\gamma_{\text{p}}\approx 5.5$\,MHz for p-polarised light and $\gamma_{\text{s}}\approx 0.975\,$MHz for s-polarised light. The lock of both polarisation modes was engaged in a two-step hierarchical process using a digital data-acquisition system (DAQ) similar to the one of the large-scale gravitational-wave detector Virgo located in Italy \cite{aVirgo2015}. First, the cavity length was controlled to match the resonance condition for p-polarised light by using PDH lock and feeding back the error signal to the piezo-electric transducer connected to the ETM, see Fig.~\ref{fig:triCav_full-scheme}. The lowest piezo resonance was observed around 10\,kHz and limited the achievable unity-gain frequency (UGF) to maximum $\sim 2-3\,$kHz. Second, the control loop for the s-polarised light was closed, controlling the AOM frequency to keep s-polarised light resonant in the cavity. The digital control filter of the s-polarisation mode was designed so that both polarisation modes exhibit a similar open-loop transfer function. Since the piezo motion is common for both polarisations, the control signal of the p-polarised light couples to the s-polarisation lock. 

The common coupling of the piezo actuation to the error signals of the two polarisation modes was used to inject a signal $\delta L(\Omega)$ in a frequency region $\sim 15-500\,$kHz far above the UGF of both control loops. The signal injection was done via a capacitor acting as high-pass filter at ~20\,kHz to prevent the signal injection from exciting piezo resonances that could peak through unity gain, and to protect the signal generator from large feedback signals at low frequencies below the UGF. 

The frequency response measurement was performed using a MokuLab signal analyser configured for swept-sine excitation. Separate response functions were recorded simultaneously for the p- and s-polarised beams using the demodulated signals of the photodetectors in reflection. The full measurement range from 15\,kHz to 500\,kHz was covered by two overlapping frequency bands to optimise the resolution and dynamic range by choosing a higher excitation amplitude for frequencies above 100\,kHz, where the response of the piezo-electric transducer was observed to degrade with increasing frequency. This frequency band covers the relevant regime in which the differential signal should experience a speed-like behaviour.

\subsection{Data analysis and results}
\label{sec:result}
The measured frequency-domain data were imported to Matlab from comma-separated values (CSV) files. Each dataset contained the excitation frequency, the magnitude response in decibels (dB), and the phase in degrees (deg). These were converted to complex transfer functions for each polarisation channel $k$ as follows: 
\begin{align}
    H_{k}(f)=10^{\frac{m_{k}(f)}{20}}\cdot e^{\mathrm{i}\phi_{k}(f)}\,,
\end{align}
where $m_{k}(f)$ is the measured gain in dB and $\phi_{k}(f)$ is the phase in radian. This conversion yields a frequency-dependent complex response $H_{k}(f)$ for each polarisation capturing both amplitude and phase behaviour of the system. Additionally, the data were sorted by frequency and concatenated into a single frequency vector with corresponding complex transfer functions for both p- and s-polarisation, denoted $H_{\text{p}}(f)$ and $H_{\text{s}}(f)$, respectively. 

\begin{figure}[h]
    \centering
    \includegraphics[width=\linewidth]{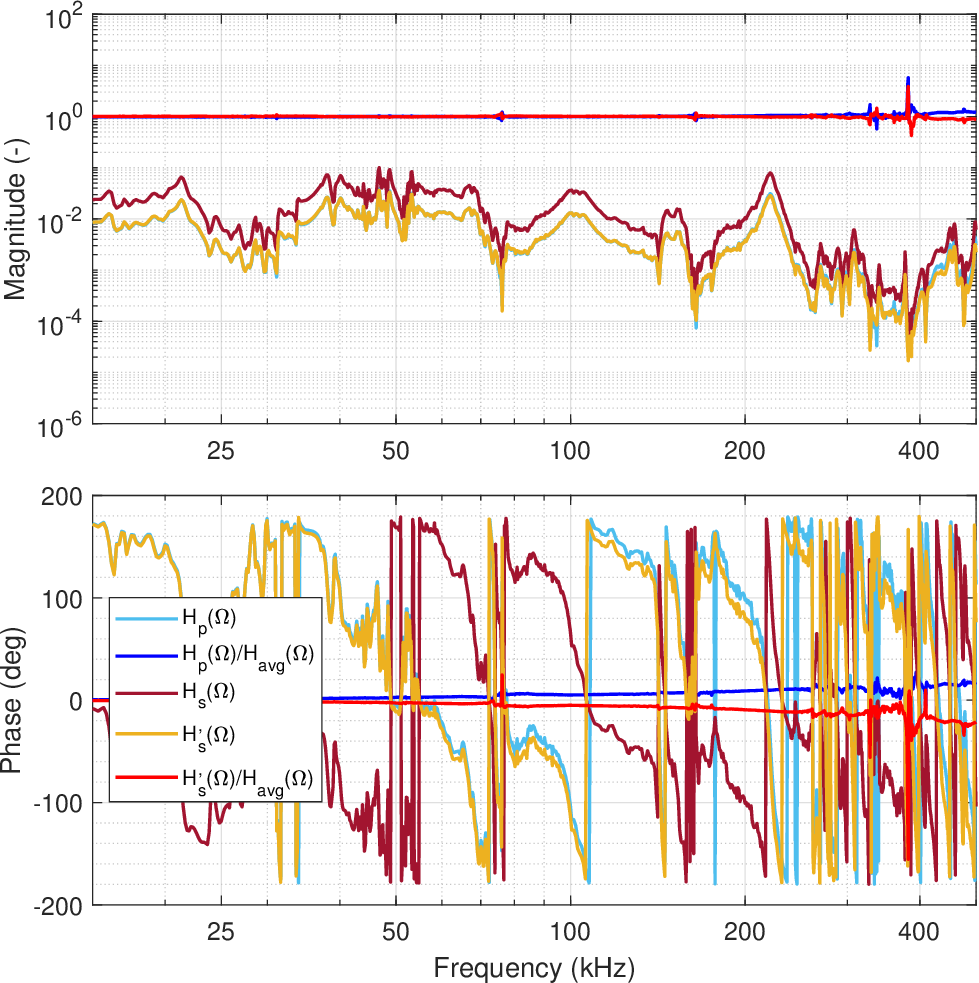}
    \caption{Raw measured complex frequency response separated into magnitude (top) and phase (bottom) for p-polarisation $H_{\text{p}}(f)$ (light blue) and s-polarisation $H_{\text{s}}(f)$ (dark red). After rescaling the s-polarisation response $H'_{\text{s}}(f)=\alpha H_{\text{s}}(f)$ (orange), the responses of both polarisations overlap. Division by the common mode response $H_{k}(f)/H_{\text{avg}}(f)$ leads to a more or less flat unity response for both polarisation modes (blue, red) as expected within the cavity bandwidth.}
    \label{fig:speed_raw-data}
\end{figure}
Since the absolute calibration between the two polarisations may differ due to gain variations, optical path differences, or photodetector responsivity, a real-valued rescaling factor $\alpha$ was estimated to align $H_{\text{s}}(f)$ with $H_{\text{p}}(f)$ using least-squared minimization over all frequency points,
\begin{align}
    \alpha = \frac{\mathbf{H}_{\text{p}}^{\dagger}\mathbf{H}_{\text{s}}}{\mathbf{H}_{\text{s}}^{\dagger}\mathbf{H}_{\text{s}}}\,,
\end{align}
with $\mathbf{H}_{k}^{\dagger}$ denoting the complex conjugate transpose of the complex response vector. The rescaled s-polarisation response is then defined as $H'_{\text{s}}(f)=\alpha H_{\text{s}}(f)$. This step minimises the integrated squared deviation between the two channels and effectively removes any constant gain mismatch. Fig.~\ref{fig:speed_raw-data} shows the raw measured response for p-polarisation (light blue) and s-polarisation (dark red) corresponding to $H_{\text{p}}(f)$ and $H_{\text{s}}(f)$ as well as the rescaled s-polarisation response $H'_{\text{s}}(f)$ (orange). 

Ultimately, we are interested in the optical response of the cavity undisturbed by resonances from the piezo-electric transducer and its clamping mechanism. To quantify the degree to which the optical system responds differently to the two polarisations, we define the common-mode response as
\begin{align}
    H_{\text{avg}}(f)=\frac{1}{2}\left[H_{\text{p}}(f)+H'_{\text{s}}(f)\right]\,,
\end{align}
and the complex fractional residual as 
\begin{align}
    H_{\text{res}}(f)=\frac{H_{\text{p}}(f)-H'_{\text{s}}(f)}{H_{\text{avg}}(f)}\,.
\end{align}
The residual function represents the polarisation-induced asymmetry in the transfer function, normalised to the average system response. Its magnitude and phase provide frequency-resolved insight into how differently the system acts on the two polarisation states. Fig.~\ref{fig:speed_raw-data} shows the common-mode response $H_{k}(f)/H_{\text{avg}}(f)$ for both p-polarisation (blue) and s-polarisation (red), which is apart from some small features close to the unity response expected for an optical cavity for signal frequencies well within its optical bandwidth.

\begin{figure}[h]
    \centering
    \includegraphics[width=\linewidth]{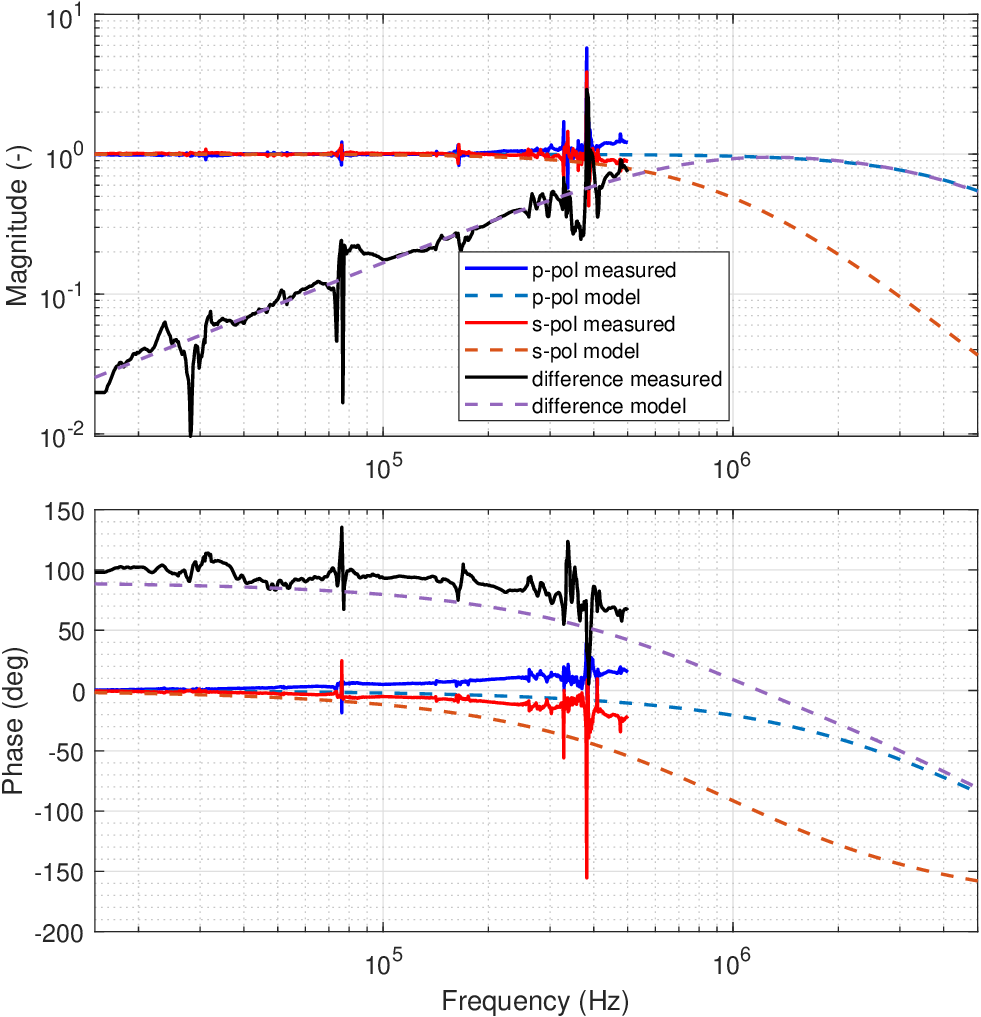}
    \caption{Comparison of measured frequency response of individual polarisation modes (blue, red) with analytical model of optical cavity response (dashed light blue, dashed orange), and differential response from post-processing of measurement data (black) with corresponding model (dashed purple).}
    \label{fig:speed_meas-vs-model}
\end{figure}
To interpret the observed response and compare against the expected speed-like behaviour, we constructed analytical transfer function models using a zero-pole-gain (ZPK) representation of the system components. Polarisation-dependent cavity responses were modeled as second-order low-pass systems with distinct cutoff frequencies given by the different bandwidths for both polarisations. These simple models provided a reference for comparing measured and predicted behavior of the system. Fig.~\ref{fig:speed_meas-vs-model} shows an overview of the comparison between model and measurement. The common mode response $H_{k}(f)/H_{\text{avg}}(f)$ for p-polarisation (blue) and s-polarisation (red), respectively, is the same as in Fig.~\ref{fig:speed_raw-data} but now compared against the model of the optical response shown in dashed lines. The measurement was performed up to 500\,kHz limited by the degrading piezo response at higher frequencies. Above 300\,kHz, the measured response is deteriorating and coherence with the excitation is reduced, therefore, the cavity pole could not be resolved for either of the polarisations. Below the cavity pole of the s-polarisation mode around 975\,kHz, the differential mode is expected to behave linearly with frequency, and the measured differential response (black) is in good agreement with the response predicted by the model (dashed purple). With this, we demonstrate the speed measurement characteristic of the EPR Speedmeter proof-of-concept experiment.

There are some dips in the measured differential response (black) coinciding with structures in the measured common-mode responses (blue, red). A potential explanation is the capacitive behaviour of the piezo-electric transducer which leads to electro-magnetic radiation emitted at certain frequencies characteristic for the piezo at hand. At these frequencies, energy from the excitation is not converted to mirror motion, hence, there is loss of coherence between optical response and excitation.

\section{Summary and discussion}
\label{sec:discussion}
In this paper, we have presented a table-top proof-of-principle experiment of the EPR Speedmeter, a concept proposed to mitigate back-action noise in gravitational-wave detectors. The EPR Speedmeter features an optical resonator for two polarisation modes with different bandwidths, which was realised in the proof-of-principle experiment by a triangular cavity with different input-mirror transmissivities for the two polarisations due to non-normal incidence. Readout of the differential response was realised by subtraction of demodulated, gain-matched reflection signals of the impedance-matched cavity with bandwidth 5.5\,MHz for p-polarisation and bandwidth 975\,kHz for s-polarisation. The measured differential response to motion of the cavity end mirror driven by a piezo-electric transducer in the frequency range 15\,kHz to 500\,kHz shows good agreement with the theoretical prediction featuring vanishing response towards DC and increase of response $\propto \mathrm{i}\Omega$ up to the cavity pole of the s-polarisation mode. 

The current setup could still be improved, for example, by optimising excitation parameters and longer averaging to yield a smoother speed response. Another aspect to investigate would be a speed readout in realtime by implementing an analogue electronics circuit for gain matching and signal subtraction. However, this would not change the main outcome of the setup, which was to demonstrate that this optical system indeed realises a speedmeter with a classical speed-like response to mirror motion. In order to move further towards technical readiness of the speedmeter for implementation as upgrade of future gravitational-wave interferometers, one would have to demonstrate back-action noise suppression experimentally. This requires a quantum-noise limited system with suspended mirrors in vacuum. 

A natural place to test the EPR Speedmeter configuration are prototype facilities such as the AEI 10\,m Prototype in Hannover (Germany), which was designed specifically to test techniques to surpass the SQL~\cite{Gossler2010}, or ETpathfinder in Maastricht (Netherlands), which is a cryogenic interferometer under construction with the aim to test technologies for the low-frequency part of the Einstein Telescope~\cite{ETpf2022}. Practical implications of implementing the EPR Speedmeter configuration and testing its alleged compatibility with established quantum-noise reduction techniques like squeezing experimentally in such a prototype facility will be subject to future work.

\section*{Acknowledgements}
The authors would like to thank David Wu and Jean-Pierre Zendri for useful discussions. This work was supported by the European Research Council (Advanced Grant SPEED 101019978).

\bibliography{bibliography.bib}

\end{document}